\documentclass[aps,prl,twocolumn,groupedaddress,showpacs]{revtex4-1}

\usepackage{graphicx}
\usepackage{amsmath}
\begin{document}


\title{Non-negligible magnetic dipole scattering from metallic nanowire for ultrasensitive deflection sensing}
\author{Zheng Xi}
\email[z.xi@tudelft.nl]{}
\affiliation{Optics Reseach Group, Delft University of Technology}
\author{H.P. Urbach}
\affiliation{Optics Reseach Group, Delft University of Technology}


\date{\today}

\begin{abstract}
It is generally believed that when a single‘ metallic nanowire is sufficiently small, it scatters like a point electric dipole. We show theoretically when a metallic nanowire is placed inside specially designed beams, the non-negligible magnetic dipole contribution along with the electric dipole resonance can lead to unidirectional scattering in the far-field, fulfilling Kerker's condition. Remarkably, this far-field unidirectional scattering encodes information that is highly dependent on the nanowire's deflection at a scale much smaller than the wavelength. The special role of small but non-negligible magnetic response and plasmonic resonance are highlighted for this extreme sensitivity as compared with the dielectric counterpart. Effects such as scattering efficiency and shape of the nanowire's cross section are also discussed.
\end{abstract}
\maketitle
The scattering of light by nanoparticles has attracted significant attention from a fundamental point of view\cite{quinten2010optical,bohren2008absorption,van1981light}. Especially, it is interesting to study the interaction of specially designed ''magnetic atoms'' with the often neglected magnetic field component of light\cite{burresi2009probing,vynck2009all,vignolini2010magnetic,karaveli2011spectral,kihm2011bethe,asenjo2012magnetic,lee2012role,rotenberg2012plasmon,yi2012diffraction,hein2013tailoring,rotenberg2013magnetic,aigouy2014mapping,coenen2014directional,rotenberg2014mapping,kasperczyk2015excitation}. This magnetic interaction has led to many fascinating phenomena such as superlensing, cloaking, negative refraction and directional scattering. However, for a simple metallic nanostructure like a nanosphere or a nanowire much smaller than the wavelength, this magnetic response is several orders of magnitude weaker, while its electric counterpart still shows large enhancement due to Localized Plasmonic Resonance(LPR). Because of this, the scattering of light as an electromagnetic wave becomes the scattering of electric field by an electric dipole instead.  

Despite of its simple geometry, the scattering by a nanowire still shows its important role in fields such as optomechanics, atomic force microscopy and quantum mechanical measurement\cite{gil2010nanomechanical,sanii2010high,gloppe2014bidimensional,ma2016sharp,de2017universal,RN196}. The far-field optical deflection measurement of the nanowire based on scattering has greatly enhanced our ability to investigate imaging and dynamics at length scales much smaller than the wavelength. The nanowire, in this case, is treated as a non-resonant electric dipole scatterer, with symmetric far-field scattering pattern\cite{sanii2010high}. The resulting change in the far-field for small deflections, is not high, which sets a limiting factor on the sensitivity of the scheme.

In this Letter, we report the non-negligible role of magnetic response of metallic nanowire much smaller than the wavelength, especially at LPR frequency, and develop a method to measure the nanowire's deflection with ultrahigh sensitivity. Specifically, we consider the scattering of a metallic nanowire inside specially designed beams rigorously using Mie theory. By moving the metallic nanowire around the point where the magnetic field is maxima with minimum electric field, we found that directional far-field scattering can be achieved even for a single metallic nanowire, fulfilling Kerker's condition\cite{kerker1983electromagnetic}. The highly directional far-field scattering encodes information about the nanowire's deflection with extreme sensitivity\cite{neugebauer2016polarization,xi2016accurate}. The enhanced sensitivity compared with the high refractive index (HRI) material counterpart is discussed in detail. Practical aspects for the scattered power and the shape of the cross section of the nanowire are also investigated.

\begin{figure}
\includegraphics[width=\linewidth]{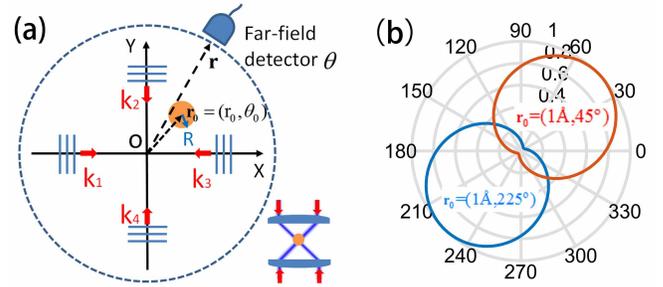}
\caption{(Color online) (a)  Geometry of the system. The nanowire of radius R (gold cross section) is displaced from the origin O by ${{\mathbf{r}}_{\mathbf{0}}}\text{=(}{{\text{r}}_{\text{0}}}\text{,}{{\text{ }\!\!\theta\!\!\text{ }}_{\text{0}}}\text{)}$. Four incoming beams $k_1$ to $k_4$ intersect at O with constructive interference  for Hz at this point. The far-field detector at $\mathbf{r}$ collects scattered far-field power along $\theta$ direction. The bottom picture shows the schematic implementation with 4Pi illumination with cylindrical lenses. (b) Far-field scattering power pattern in XY plane when the deflection is ${{\mathbf{r}}_{\mathbf{0}}}\text{=(1\AA}\text{,4}{{\text{5}}^{\text{o}}}\text{)}$ (Red curve) and  ${{\mathbf{r}}_{\mathbf{0}}}\text{=(1\AA}\text{,22}{{\text{5}}^{\text{o}}}\text{)}$(Blue curve).}
\end{figure}

The considered geometry is sketched in Fig.1(a). A silver nanowire of radius R=10nm (only the cross section is shown) is placed within four orthogonally oriented TE beams of equal intensity (the magnetic field is polarized along the z-axis).  The center position of the nanowire is described by ${{\mathbf{r}}_{\mathbf{0}}}=({{r}_{0}},{{\theta }_{0}})$ in polar coordinates. The phases of the four beams are chosen such that the Hz component is maximum at the origin. This geometry has been extensively used to trap and manipulate atoms in an optical lattice\cite{hemmerich1992radiation,albaladejo2009scattering,barry2014magneto} and can be implemented using common dark-field illumination scheme shown at the bottom of Fig. 1(a). 

The silver nanowire is placed in vacuum and it exhibits a localized plasmonic resonance at $\lambda =337nm$\cite{palik1998handbook} under quasi-static approximation\cite{bohren2008absorption}. Fig.1(b) shows the far-field scattering pattern at this wavelength when the nanowire is displaced from the center origin O by ${{\mathbf{r}}_{\mathbf{0}}}\text{=(1\AA}\text{,4}{{\text{5}}^{\text{o}}}\text{)}\ \text{and}\ {{\mathbf{r}}_{\mathbf{0}}}\text{=(1\AA}\text{,22}{{\text{5}}^{\text{o}}}\text{)}$. The silver nanowire, in this case, no longer scatters like an electric dipole, but with highly asymmetric pattern when it is away from the origin O. The scattered power always reaches the maximum along deflection direction $\theta ={{\theta }_{0}}$ and has minimum in the opposite direction $\theta ={{\theta }_{0}}+\pi$. For opposite deflection directions, the maxima and minima switches, even for ${{r}_{0}}=\text{1\AA}$.

In order to understand this phenomenon, a theoretical model  based on Mie theory to calculate the far-field scattering at the detector $\mathbf{r}$ is developed. In the original Mie scattering theory, only one incident plane wave is considered and the corresponding far-field can be calculated as\cite{van1981light}:
\begin{equation}
{{H}_{z,mie}{(\mathbf{r})}}=-4i{{G}_{\infty }}[{{a}_{0}}+2\sum\limits_{n=1}^{\infty }{{{a}_{n}}\cos(n\theta )}] 
\end{equation}
where the terms ${{a}_{n}}$ are the Mie coefficients which contain information of the scatterer, with ${{a}_{0}}$, ${{a}_{1}}$ and ${{a}_{2}}$ describing the induced magnetic dipole moment, electric dipole moment and electric quadrupole moment respectively and other ${{a}_{n}}$ for higher order multipole terms\cite{kallos2012resonance}. The scalar 2D Green's function ${{G}_{\infty }}=\frac{i}{4}H_{0}^{(1)}({{k}_{0}}r)$ is used, where $H_{0}^{(1)}({{k}_{0}}r)$ is the zeroth order Hankel function of the first kind and ${{k}_{0}}$ is the wave vector in free space. 
In our case, rotations for four beams are required and the final scattering field can be added, which is:
\begin{equation}
{{H}_{z,far}{(\mathbf{r})}}=-\sum\limits_{m=1}^{4}{4i{{G}_{\infty }}\{{{a}_{0}}+2\sum\limits_{n=1}^{\infty }{{{a}_{n}}\cos [n(\theta +{{\theta }_{m}})]\}{{e}^{i{{\mathbf{k}}_{\mathbf{m}}}
\cdot \mathbf{r}}}}}
\end{equation}
$\theta_m$ is the relative incident angle of the four beams and the term ${{e}^{i{{\mathbf{k}}_{\mathbf{m}}}\cdot \mathbf{r}}}$ is due to extra phase gained by displacement $\mathbf{r}=({{r}_{0}},{{\theta }_{0}})$ for different beams. In the calculation, we fix the origin of the nanowire at the origin of the coordinate system and treat the displacement as extra phases for each incident beam. Keeping the first three ${{a}_{n}}$ terms, the scattered far-field Hz along $\theta$ direction is:
\begin{widetext}
\begin{equation}
\begin{aligned}
  {{H}_{z,far}}(\theta )=-4i&{{G}_{0}}[2{{a}_{0}}\cos ({{k}_{0}}{{r}_{0}}\cos {{\theta }_{0}})+4i{{a}_{1}}\cos \theta \sin ({{k}_{0}}{{r}_{0}}\cos {{\theta }_{0}}) 
  +4{{a}_{2}}\cos (2\theta )\cos ({{k}_{0}}{{r}_{0}}\cos {{\theta }_{0}}) \\ 
 &\ +2{{a}_{0}}\cos ({{k}_{0}}{{r}_{0}}\sin {{\theta }_{0}})+4i{{a}_{1}}\sin \theta \sin ({{k}_{0}}{{r}_{0}}\sin {{\theta }_{0}})  
  -4{{a}_{2}}\cos (2\theta )\cos ({{k}_{0}}{{r}_{0}}\sin {{\theta }_{0}})]  
\end{aligned}
\end{equation}
\end{widetext}
For displacement ${{r}_{0}}$ much smaller than wavelength, i.e. ${{k}_{0}}{{r}_{0}}<<1$, we arrive at the final expression for the scattered far-field:
\begin{equation}
{{H}_{z,far}}(\theta )=-16i{{G}_{\infty }}[{{a}_{0}}+i{{a}_{1}}{{k}_{0}}{{r}_{0}}\cos (\theta -{{\theta }_{0}})]
\end{equation}

Looking at this equation, the far-field contribution from the electric dipole ${{a}_{1}}$ is reduced because ${{k}_{0}}{{r}_{0}}<<1$,  thus the contribution of the magnetic dipole ${{a}_{0}}$ becomes more important. It is interesting to note that although the first three Mie coefficients are considered, only the first two dipolar terms play important roles here, the term ${{a}_{2}}$ which corresponds to the contribution of the electric quadrupole cancels out in the above expression. This would allow the current analysis to be applied to larger radius as well.
In the direction of $\theta ={{\theta }_{0}}$ and $\theta ={{\theta }_{0}}+\pi$, the cosine term equals 1 and -1 such that the far-fields in these two directions take the value: 
\begin{equation}
\begin{aligned}
  & {{H}_{z,far}}(\theta ={{\theta }_{0}})=-16i{{G}_{\infty }}({{a}_{0}}+i{{a}_{1}}{{k}_{0}}{{r}_{0}}) \\ 
 & {{H}_{z,far}}(\theta ={{\theta }_{0}}+\pi )=-16i{{G}_{\infty }}({{a}_{0}}-i{{a}_{1}}{{k}_{0}}{{r}_{0}})  
\end{aligned}
\end{equation}

If the nanowire can be designed that for a certain displacement ${{r}_{0}}$, ${{a}_{0}}=i{{a}_{1}}{{k}_{0}}{{r}_{0}}$, then 
\begin{equation}
\begin{aligned}
  & {{H}_{z,far}}(\theta ={{\theta }_{0}})=-16i{{G}_{\infty }}\cdot 2{{a}_{1}}{{k}_{0}}{{r}_{0}} \\ 
 & {{H}_{z,far}}(\theta ={{\theta }_{0}}+\pi )=0 \\ 
\end{aligned}
\end{equation}

In this case, the field in one direction is zero and the other one is the maximum. A strong asymmetry is presented in the far-field as shown in Fig. 1(b).  The expression
\begin{equation}
{{a}_{0}}=i{{a}_{1}}{{k}_{0}}{{r}_{0}}            
\end{equation}
gives the condition for the maximum asymmetry in the far-field. 

Generally, the calculation of ${{a}_{0}}$ and ${{a}_{1}}$ involves Bessel functions and Hankel functions which lacks a clear physical interpretation. However, when the radius R of the nanowire is much smaller than the wavelength of light, the following approximations can be made\cite{quinten2010optical}:
\begin{equation}
\begin{aligned}
&{{a}_{0}}\approx -i\pi k_{0}^{4}{{R}^{4}}\frac{\epsilon -1}{32} \\
&{{a}_{1}}\approx -i\pi k_{0}^{2}{{R}^{2}}\frac{\epsilon -1}{4(\epsilon +1)}
\end{aligned}
\end{equation}
with their ratio:
\begin{equation}
\frac{{{a}_{1}}}{{{a}_{0}}}\approx \frac{8}{k_{0}^{2}{{R}^{2}}(\epsilon +1)}
\end{equation}
$\epsilon =\epsilon '+i\epsilon ''$ is the permittivity of the nanowire's material. For infinitesimal metallic nanowire, the contribution from the magnetic dipole moment ${{a}_{0}}$ is considered to be a high order term ($k_{0}^{4}{{R}^{4}}$) as compared with the electric dipole ${{a}_{1}}$ under plane wave excitation, and thus is generally neglected. The far-field scattering resembles the one produced by the electric dipole. However, the denominator in ${{a}_{1}}$ indicates a localized plasmonic resonance when ${\epsilon }'=-1$, or more generally a Fr\"ohlich resonance\cite{quinten2010optical}. At this wavelength, the ratio between ${{a}_{1}}$ and ${{a}_{0}}$ becomes purely imaginary and this fulfills the requirement by Eq.  (7), which leads to:
\begin{equation}
{{r}_{0}}=\frac{\pi {{R}^{2}}\epsilon _{p}^{''}}{4{{\lambda }_{p}}}
\end{equation}
where $\lambda_p$ is the LPR wavelength and $\epsilon _{p}^{''}$ is the imaginary part of the permittivity at resonance. A sharp asymmetry in the far-field can be observed when the nanowire is at position ${{r}_{0}}$. This asymmetric far-field pattern can't be produced by electric dipole alone, it can only be observed when the non-negligible magnetic contribution from the metallic nanowire is taken into account and optimized appropriately. Considering the value for the silver nanowire, ${{\lambda }_{p}}\approx 337nm,\ R=10nm\ $and $\epsilon _{p}^{''}\approx 0.58$\cite{palik1998handbook}, a displacement of ${{r}_{0}}\approx 1\text{\AA}$ can lead to strong asymmetry in the far-field, which explains the extreme sensitivity in Fig. 1(b). This rigorous treatment provides an accurate model for deducing very small deflection ${{r}_{0}}$ of the nanowire by measuring the asymmetry in the far-field. Inversely, it also serves as a design guideline for choosing the nanowire's property for optimum sensing for a certain deflection ${{r}_{0}}$: for the measurement of smaller ${{r}_{0}}$, a smaller nanowire radius R, a smaller imaginary part $\epsilon _{p}^{''}$  and a longer resonance wavelength $\lambda_p$ are preferred. What is interesting to note here is the inverse proportionality to the resonance wavelength $\lambda_p$.  For a recently highlighted material indium tin oxide\cite{alam2016large}, the resonance is at ${{\lambda }_{p}}=1.42\mu m$ with $\epsilon _{p}^{''}\approx 0.5$ similar to silver. According to Eq. (10), a radius the same as the silver case leads to ${{r}_{0}}$ that is about 4 times smaller, but at infrared wavelength. The general condition for Fr\"ohlich resonance allows us to explore materials involving surface phonon resonance such as SiC\cite{hillenbrand2002phonon}. The highly tunable plasmonic resonance in heavily doped semiconductor or core-shell materials also enriches the flexibility of the present scheme\cite{quinten2010optical}. 
\begin{figure}
\includegraphics[width=\linewidth]{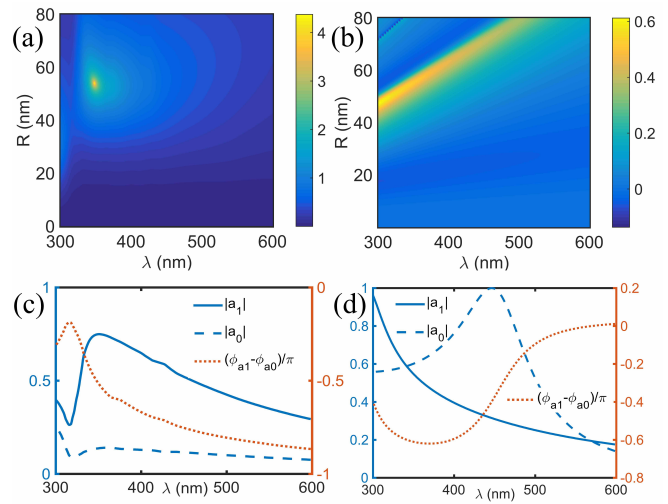}
\caption{(Color online)(a) Plot of sensitivity S for silver nanowire for a displacement of 10nm. (b) The same as (a) but for HRI dielectric nanowire. (c) The contribution from electric dipole coefficient ${{a}_{1}}$, magnetic dipole coefficient ${{a}_{0}}$ and their phase difference for silver nanowire of radius R=54nm. (d) The same as (c) but for HRI dielectric nanowire of radius R=47nm.}
\end{figure}

We would like to briefly discuss the case for a perfect electric conductor (PEC) nanowire and a pure dielectric nanowire (real $\epsilon$). For PEC ($\epsilon =-\infty $), the coefficients ${{a}_{0}}$ and ${{a}_{1}}$ are out of phase. For the pure dielectric case, ${{a}_{0}}$ and ${{a}_{1}}$ are in phase. The lack of $\pi /2$ phase difference as required by Eq. (7) in these two cases largely prevents the strong asymmetric far-field scattering.    

Although the above analysis is based on quasi-static approximation, the condition for maximum asymmetry given by Eq. (7) is rigorous for a larger radius R, as long as only the first three Mie coefficients are dominate. A closer look at Eq. (7) would reveal the general condition for maximum asymmetry at ${{r}_{0}}$: (1) The  electric dipole contribution ${{a}_{1}}$ should be much larger than the magnetic dipole contribution ${{a}_{0}}$ for ${{k}_{0}}{{r}_{0}}<<1$ and (2) there should be a $\pi/2$ phase difference between them. This happens near the plasmonic resonance for metal or near the electric dipole resonance for HRI dielectric materials. A very strong asymmetry can also be expected around the \emph{magnetic anapole mode} where ${{a}_{0}}$ is nearly zero\cite{wei2016excitation}. It is important to emphasize the importance of the magnetic dipole contribution ${{a}_{0}}$ here, because without it, according to Eq. (5), the far-field would be symmetric.
 
To study this effect in more detail, we optimize the radius and working wavelength of the nanowire to achieve maximum sensitivity for a displacement measurement of 10nm using the above rigorous treatment. We take the logarithmic power ratio in opposite directions 
\begin{equation}
S={{\log }_{10}}|\frac{H{_{z,far}}(\theta ={{\theta }_{0}}+\pi )}{H{_{z,far}}(\theta ={{\theta }_{0}})}{{|}^{2}}
\end{equation}
to quantify the sensitivity. The results are shown in Fig. 2(a) and Fig. 2(b) for the case of a silver nanowire and a HRI dielectric nanowire with refractive index n=3.5.   

A clear difference can be seen between the two cases: for silver, a very large asymmetry of S=4.4 can be observed at 348nm for the radius R=54nm, whereas for the HRI dielectric case, only moderate asymmetry of S=0.6 can be observed at 300nm when R=47nm.
		
In order to explain the difference between these two cases, we plot the absolute value of ${{a}_{0}}$, ${{a}_{1}}$ and their phase difference in Fig.2 (c) and (d), corresponding to the optimized radius R achieving maximum S. We are interested in the wavelength where the two coefficients have $\pi /2$ phase difference as required by Eq. (7). For silver, this happens at 348nm, which corresponds to the LPR as indicated by a peak in ${|{a}_{1}|}$. Because there is no magnetic resonance for a silver nanowire, the ratio ${|{a}_{1}|}/{|{a}_{0}|}$ can be large, about 5 times in this case. Substituting this into Eq. (7) would yield the displacement with highest sensitivity at ${{r}_{0}}=10nm$, which explains why S is so high at this point in Fig.2 (a). For the HRI dielectric case, in contrast with silver, there is an additional magnetic resonance for ${|{a}_{0}|}$ peaked at 435nm besides the electric dipole resonance for ${|{a}_{1}|}$  peaked at 292nm (not shown). Because of these two resonances, there are two wavelengths at which the two coefficients have $\pi /2$ phase difference, namely at 311nm and 428nm respectively. Only the first wavelength is of interest because we are looking for a large $|{{a}_{1}|}/{|{a}_{0}|}$. However, at this wavelength, the tail of the ${|{a}_{0}|}$ resonance is still high and ${|{a}_{1}|}/{|{a}_{0}|}$  hereby decreased to about 1.4. Therefore, S is only 0.6, which is much lower than the metal case. It is expected that a large asymmetry occurs at a displacement ${{r}_{0}}$ of 34nm to match ${|{a}_{1}|}/{|{a}_{0}|}$ in the HRI nanowire. It is interesting to note that if we move the nanowire around the  maxima of the electric field E, then the second wavelength can be used\cite{neugebauer2016polarization}. However, as pointed above, the strong overlapping between the two resonances decreases the ratio between the two coefficients and thus, sets a limit to the minimum ${{r}_{0}}$ that can be sensed using HRI for strong asymmetrical scattering compared with the metal case\cite{neugebauer2016polarization}. 
		
Next, we would like to discuss two aspects which are important for implementing this scheme into practice: namely the amount of the scattered power and the shape of the nanowire.
For the real situation, there is no perfect plane wave. The definition of scattering cross section is not valid anymore for an inhomogeneous illumination. To quantitatively describe the scattering ability of the nanowire, we introduce the scattering efficiency 
\begin{equation}
\kappa =\frac{{{P}_{scat}}}{{{P}_{inc}}} 
\end{equation}
defined as the ratio between the total scattered power ${{P}_{scat}}$ to the total input power of the four beams ${{P}_{inc}}$. It is interesting to study the case of the four focused incident beams because it introduces higher magnetic or electric energy density around the origin O. Going back to our initial configuration in Fig. 1(a), the direction of asymmetric scattering only depends on the direction of the deflection $\theta_0$. If we rotate the four beams clockwisely or counter-clockwisely, the maximum asymmetry still appears along $\theta=\theta_0$. Because the focused beam can be decomposed into plane waves, the scattering of the four focused beams would yield similar asymmetry in the far-field. In the focused beam case, the scattering power can be further increased by using higher numerical aperture.

In Fig. 3(a), we plot ${{\log }_{10}}\kappa $ as a function of relative displacement ${{r}_{0}}/{{r}_{0max}}$ along the Y axis shown in Fig. 1(a).  We optimized the radius of the silver nanowire to be R=70nm, 30nm and 10nm, such that the sensitivity S is 1 at the displacement of ${{r}_{0max}}=$10nm, 1nm and $\text{1\AA}$ respectively. The N.A. is chosen to be N.A.=0.9 and N.A.=0.6 and the wavelength is fixed at 337nm.
\begin{figure}
\includegraphics[width=\linewidth]{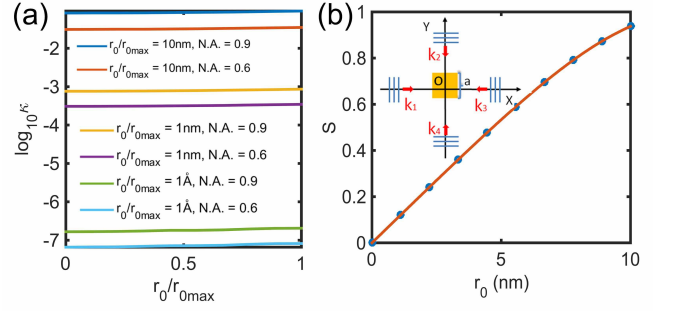}
\caption{(Color online)(a) Scattering efficiency for optimized radius for different displacement sensing. The maximum S=1 at  ${{r}_{0}}/{{r}_{0max}}$ is obtained for all cases. (b) The sensitivity S along Y axis for a nanowire of square cross section with side length a = 70nm at the wavelength of 447nm obtained by retrieving ${{a}_{0}}$ and ${{a}_{1}}$ from Fourier decomposition of the scattering far-field under single plane wave excitation (red curve) and by numerical simulation using finite element method (blue dots).}
\end{figure}
			
In all three cases, using a large N.A. improves the scattering efficiency due to increased energy density around the origin. For decreasing ${{r}_{0max}}$ from 10nm to $\text{1\AA}$, the scattering ratio $\kappa$ drops from about ${{10}^{-1}}$ to ${{10}^{-7}}$, which is still detectable\cite{piliarik2014direct}. However, $\kappa$ stays nearly constant for relative displacements ${{r}_{0}}/{{r}_{0max}}$ from 0 to 1. This is because as the nanowire is displaced from origin, although the scattered power due to magnetic dipole decreases, the scattered power due to electric dipole increases and these two effects balance each other making the total scattered power nearly constant. A compromise has to be made on optimizing the radius of the nanowire to maintain detectable scattering and on achieving the highest sensitivity for far-field scattering asymmetry. Experimental schemes involving spilt photodiode or homodyne detection would yield better sensitivity even for low scattering power.
						
The above discussions assume a nanowire of a perfect circular cross section. However, in a real situation, it is not always the case. One possible substitution would be a nanowire of square cross section of length ${a}$ as shown in the inset of Fig. 3(b). Looking back to Eq. (1), it has the form of decomposing the scattered far-field into Fourier components, with ${{a}_{0}}\ \text{and}\ 2{{a}_{n}}$ the corresponding Fourier coefficients. A similar decomposition can be applied to the square cross section case for plane wave excitation to find the corresponding ${{a}_{0}}$ and ${{a}_{1}}$. Then substitute these two coefficients into Eq.(5) to calculate the scattered far-field and finally S. We next calculate S as the nanowire of square cross section moves along Y axis using Eq. (11) (red curve in Fig. 3(b) ) and compare it with finite element simulation (blue dot in Fig. 3(b) ) at the wavelength of 447nm.  A very good agreement is achieved for two cases. But the semi-analytical approach based on Eq. (11) allows faster optimization for deflection sensing.

In conclusion,  we have shown that the normally neglected magnetic response from a metallic nanowire can be used to develop an ultrasensitive method for the measurement of deflection. For radius much smaller than the wavelength, the metallic nanowire working at LPR wavelength yields the maximum sensitivity to small deflection, which is in particular much higher than that can be achieved with the pure HRI dielectric counterpart. It is expected that the rethinking of the role of magnetic dipole scattering in metallic nanowire would provide new guidelines for the development of nanowire sensors that are important in various metrology applications and also yields interesting physics involving interactions between the magnetic component of light and matter.

\begin{acknowledgments}
The authors would like to thank Lei Wei for inspiring discussions.
%
\end{acknowledgments}


%

%

\end{document}